# Astro2020 Science White Paper

# Studying the Reionization Epoch with QSO Absorption Lines

**Thematic Areas:**   ☐ Planetary Systems   ☐ Star and Planet Formation
  ☐ Formation and Evolution of Compact Objects   ☒ Cosmology and Fundamental Physics
  ☐ Stars and Stellar Evolution   ☐ Resolved Stellar Populations and their Environments
  ☒ Galaxy Evolution       ☐ Multi-Messenger Astronomy and Astrophysics

**Principal Author:**
Name: George D. Becker
Institution: University of California, Riverside
Email: george.becker@ucr.edu
Phone: +1 (951) 827-5268

**Co-authors:** Anson D'Aloisio (UC Riverside), Frederick B. Davies (UC Santa Barbara), Joseph F. Hennawi (UC Santa Barbara), Robert A. Simcoe (MIT)

**Abstract**: Absorption signatures in the spectra of QSOs are one of our most powerful tools for studying galactic and intergalactic environments at high redshifts. With the discovery of QSOs out to z > 7, QSO absorption lines are now tracing the end stages of reionization on multiple fronts using the hydrogen Lyα forest and heavy element absorbers. Next-generation QSO absorption line studies with large optical/IR telescopes will reveal in detail how the first galaxies emerged form the cosmic web, transformed their circum- and inter-galactic environments, and completed the last major phase transition of the Universe. These efforts will complement other *upcoming* studies of reionization, such as those with *JWST,* ALMA, and redshifted 21cm experiments.

*Facilities emphasized:* ELTs, JWST, moderate- and high-resolution optical and near-infrared spectroscopy

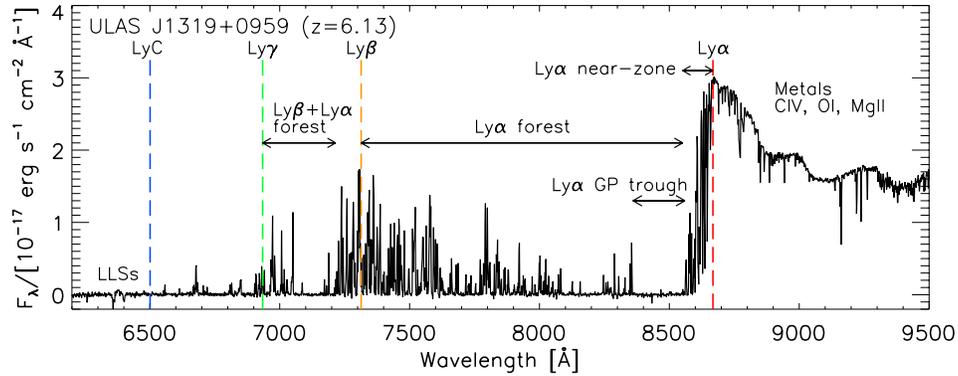

**Figure 1:** Schematic of a high-redshift QSO spectrum, here at z = 6.1. The Lyα forest traces diffuse hydrogen in the IGM, which becomes almost completely opaque by z ∼ 6. Metal absorption lines, which trace the chemically enriched gas around galaxies, can be seen redward of the Lyα emission peak. *Figure: Becker, Bolton & Lidz (2015)*

# I. Introduction: Recent Advances in Reionization

The epoch of reionization was a transformational event in cosmic history. Within one billion years after the Big Bang, radiation from the first luminous objects ionized nearly every hydrogen atom in the intergalactic medium (IGM), completing the last major global phase transition. Understanding reionization is now a primary goal of astrophysics for several reasons. First, it represents a critical epoch in the cosmic history of baryons. Second, when and how reionization occurred gives unique insight into the formation of the first galaxies, many of which are too faint to observe directly. Finally, understanding reionization is critical for fully exploiting the cosmic microwave background (CMB) and the IGM as cosmological probes.

The past several years have seen rapid advances in reionization studies, with multiple lines of evidence beginning to suggest a midpoint of reionization as late as z ∼ 7–8. This timing is supported by CMB measurements from Planck (Planck Collaboration 2018), the rapid decline from z ∼ 6 to 8 in the fraction of star-forming galaxies that show Lyα in emission (e.g., Schenker et al. 2012; Treu et al. 2013; Pentericci et al. 2014; Hoag et al. 2019; Mason et al. 2019), Lyα damping wings in the spectra of the z ∼ 7 QSOs (Mortlock et al. 2011; Davies et al. 2018b; Greig et al. 2019), and the thermal history of the IGM (Boera et al. 2019). At the same time, observations of transmitted flux in the Lyα forest of z > 6 QSOs indicate that reionization is largely complete by z ∼ 6 (e.g., Fan et al. 2006, McGreer et al. 2015).

Fundamental uncertainties nevertheless remain about when and how reionization occurred. Constraints on the timing remain broad, and it is not yet clear that the constraints are all mutually consistent. The nature of the sources also remains elusive. The number density of AGN appears too low to produce enough photons to ionize the IGM by z ∼ 6 (e.g., McGreer et al. 2018). Massive stars produce enough ionizing photons, but it is unclear whether enough of these photons escape from the galaxies in which they form. Clearly, a better understanding of how and when reionization occurred is needed to form a self-consistent picture of the growth of early galaxies.

**As reionization studies advance over the coming decade, spectroscopy of high-redshift QSOs will be one of our most powerful observational tools.** QSO absorption lines trace baryons from galactic to cosmological scales, encoding detailed information about the density, ionization, temperature, and chemical composition of the intervening gas (Figure 1).

Spectroscopy of QSOs already plays a key role in determining how the IGM and galaxies evolve near reionization. There are now substantial samples of QSOs out to z ~ 7, and many more will be discovered by *Euclid* and *WFIRST*. Below we discuss the opportunities to transform reionization studies with QSOs into a field of precision astrophysics.

## II.   Science Opportunities for QSO Absorption Lines

**Neutral Islands at the End of Reionization**

One of the most striking features of the IGM near reionization is the extreme scatter in intergalactic Lyα opacity over 5 < z < 6 (Fan et al. 2006; Becker et al. 2015; Bosman et al. 2018; Eilers et al. 2018). The Lyα forest in QSO spectra shows far more variation in the amount of transmitted flux at these redshifts than can be explained by differences in the large-scale density field alone (Becker et al. 2015). The longest (~160 comoving Mpc), most opaque known Lyα trough is associated with an underdensity of galaxies (Becker et al. 2018), suggesting that the variations in opacity require either strong fluctuations in the ionizing UV background (Davies et al. 2016; D'Aloisio et al. 2018a), or that reionization ends as late as z ~ 5–5.5 (Kulkarni et al. 2018), far later than previously thought.

The Lyα forest offers multiple probes of late reionization. Neutral patches at z < 6 will appear as long "dark gaps" in Lyα transmission, motivating measurements of the dark gap width distribution. Damping wing signatures in the forest and the presence of deuterium absorption lines may directly indicate the presence of neutral gas (Malloy & Lidz 2015). The redshifts over which the first Lyα and Lyβ transmission peaks appear, moreover, should reflect the redshift interval over which reionization ends (Chardin et al. 2018; Garaldi et al. 2019).

<u>Future Prospects</u>: Fully exploiting the Lyα forest near reionization requires moderate-resolution (R > 5,000) optical spectroscopy of enough QSOs to sample the complex topology of neutral and ionized regions. Current facilities are sensitive enough to detect transmission peaks with sufficient sensitivity in only the ~20 brightest known z > 6 QSOs. ELTs, however, can efficiently expand this sample to >100 lines of sight at z > 6 using QSOs that are already known (e.g., Bañados et al. 2016; Reed et al. 2017). Such a dataset would robustly sample the IGM during the end stages of reionization, complementing 21cm experiments such as HERA and SKA. Wide-area galaxy surveys (e.g., with *WFIRST*) in QSO fields will also help to determine how properties of the IGM correlate with environment during this epoch.

**The Thermal History of the IGM**

During reionization, ionization fronts photoheat the cold IGM to temperatures of T = 20,000–30,000 K (e.g., Shapiro & Giroux 1987, Miralda-Escudé & Rees 1994; D'Aloisio et al. 2018b). The reionized gas then cools over cosmological time scales (Δz = 1–2) by the expansion of the Universe and other well-understood processes (e.g., Hui & Gnedin 1997; Upton Sanderbeck et al. 2016). The temperature of the IGM therefore retains information about this heat injection even well after reionization ends.

A number of observational tools have been developed to extract the thermal history of the IGM from the thermal broadening of Lyman-α forest absorbers (e.g., Schaye et al. 2000; Zaldarriaga et al. 2001; McDonald et al. 2001; Lidz et al. 2010; Becker et al. 2011a). At z = 2–4, temperature measurements have successfully traced the thermal evolution of the IGM during He II reionization (e.g., Becker et al. 2011a; Boera et al. 2014; Hiss et al. 2018; Walther et al.

2019). Recently, temperature measurements have been pushed to z = 5.4, where they are sensitive to the timing and duration of hydrogen reionization (Figure 2) (Viel et al. 2013; Boera et al. 2019; Walther et al. 2019).

Perhaps even more significant is the fact that the Lyα forest contains information beyond the instantaneous gas temperature. After reionization, the IGM responds dynamically to the increased pressure, smoothing out cosmological density fluctuations on small scales. In contrast to thermal broadening, this "Jeans smoothing" depends on the time-integrated thermal history of the gas (e.g., Shapiro et al. 1994; Rorai et al. 2013; Kulkarni et al.2015). Recent studies have successfully made the first joint measurements of IGM temperature and Jeans smoothing out to z ~ 5 (Boera et al. 2019; Walther et al. 2019), exploiting the fact that the effects operate over different scales in the Lyman-α forest power spectrum (Nasir et al.2016; Oñorbe et al. 2017).

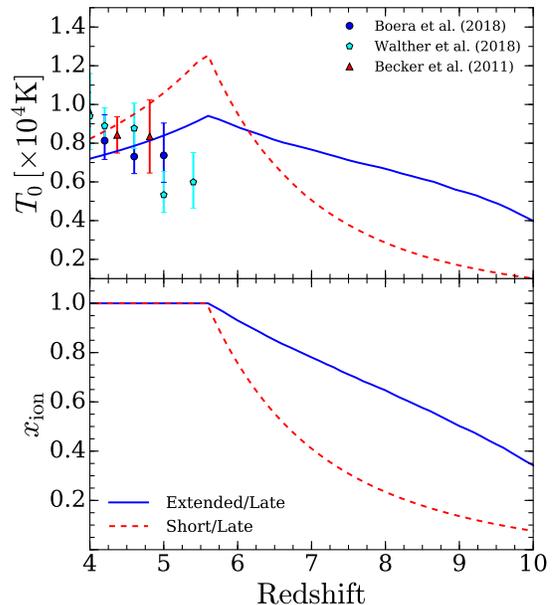

**Figure 2.** IGM temperature at mean density (top panel) and ionized hydrogen fraction (bottom panel) during and after reionization. Measurements of the IGM thermal history place strong constraints on the timing and duration of reionization. The solid and dashed lines are semi-analytic reionization models following Upton Sanderbeck et al. (2016).

Future prospects: Measurements of the IGM thermal history now constrain the timing of reionization with precision comparable to *Planck*, providing a fully independent check on the CMB results (Boera et al. 2019). The current studies are limited, however, by the small number (~15) of z > 5 QSOs that are bright enough to study at high resolution (R ~ 40,000). A high-resolution optical spectrograph on a 25–30m telescope would enable samples of >100 QSOs, delivering some of the most robust and informative constraints on reionization available.

**QSO Damping Wings**

During reionization, the substantial fraction of neutral hydrogen in the IGM leads to absorption redward of rest-frame Lyα, known as the "damping wing" (Miralda-Escude 1998). Conclusive detection of this signal in QSO spectra would be a smoking gun for reionization, and allow for measurements of the hydrogen neutral fraction $x_{HI}$ as a function of redshift. In addition, the transparent "proximity zone" in front of these QSOs constrains the duration of their UV-luminous phase, providing key insights into the accretion history of the first supermassive black holes.

Measuring $x_{HI}$ from the damping wing is not trivial, however. In order to detect the smooth damping wing absorption signal, one must be able to predict the unabsorbed QSO spectrum near the Lyα emission line. State-of-the-art methods leveraging thousands of SDSS spectra of lower redshift QSOs are now able to do this (Davies et al. 2018a, Greig et al. 2017a). Well calibrated medium-resolution near-infrared spectroscopy (e.g., from *JWST*) is crucial for accurately measuring of the rest-frame UV broad emission lines used to predict the spectrum close to Lyα. In addition, converting the damping wing signal into a constraint on $x_{HI}$ requires large-volume semi-numerical simulations with which to model the patchy reionization topology around massive QSO-hosting dark matter halos (Alvarez & Abel 2007; Lidz et al. 2007; Mesinger et al. 2011,2016; Davies et al. 2018b).

Existing damping wing analyses have examined the two highest QSOs known, ULAS J1120+0641 at z = 7.09 and ULAS J1342+0928 at z = 7.54 (Greig et al. 2017b, 2019; Davies et al. 2018b). These works demonstrate that an individual reionization-epoch QSO can constrain $x_{HI}$ to ±0.2 (Figure 3), where the precision is limited predominantly by the stochasticity of ionized bubbles during reionization.

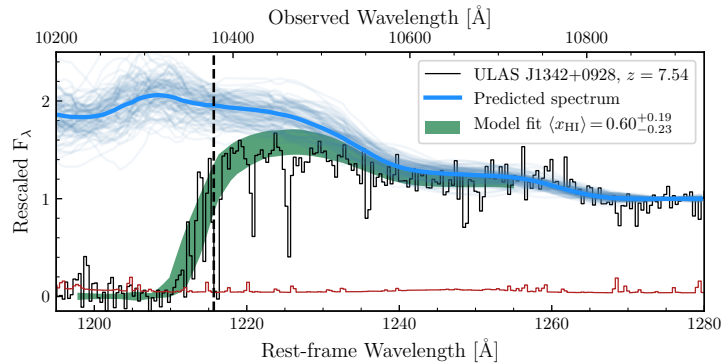

**Figure 3.** Damping wing absorption in a QSO at z = 7.5. The observed spectrum shows a deficit of flux reward of 1216 Å with respect to the predicted continuum (blue line). A fit to the damping wing absorption indicates a substantial neutral fraction in the surrounding IGM.

Future prospects: Surveys hunting for more z > 7 QSO are ongoing, and another two z ~ 7 QSOs have already been published (Wang et al. 2018; Yang et al. 2018). Future surveys are expected to reveal many more reionization-epoch QSOs within the next decade, particularly in the era of space-based near-infrared surveys with *WFIRST* and *Euclid*. Near-IR spectra of these objects with *JWST* and large ground-based facilities will make it possible to measure the neutral fraction of the IGM with <10% precision over the full history of reionization.

**Observations of Heavy Elements**

Heavy-element absorption systems at z > 6 yield sensitive measurements of galactic and circumgalactic gas, even as neutral intergalactic hydrogen absorbs all flux at wavelengths below Lyman alpha in the QSO's rest frame. The high luminosity and point-like morphology of QSOs enable quantitative physical characterization of baryonic matter beyond the reach of magnitude-limited galaxy surveys. Indeed, most galaxies responsible for reionization and early metal pollution may be too faint to observe even with *JWST* (Finlator et al 2013), yet they can be studied via their absorption signatures.

Infrared echelle spectra have recently uncovered robust populations of metal absorbers out to nearly z = 7 (Bosman et al. 2017; Chen et al 2017). These systems display a rich array of atomic transitions, including α- and Fe-group species with low ionization energies (E ~ 1 Ryd) such as Mg II, Si II, C II, O I, Fe II, and Al II, as well as benchmark species with higher ionization energies (E ~ 3-4 Ryd) such as C IV and Si IV. Low-ionization lines typically track the metallicity of circum-galactic gas, while high-ionization lines help to resolve degeneracies between abundance and ionization (e.g., Glidden et al 2016).

Exploratory surveys at z > 6 have begun to reveal distinct trends in the evolution of different ions. The comoving linear density of low-ionization Mg II remains statistically constant from z < 0.5 to z > 6.5 (Chen et al 2017), while the detection rate of highly-ionized gas at z > 6 traced by C IV declines by more than an order of magnitude from its peak at z ~ 3 (D'Odorico et al 2013). Approaching reionization, the classic picture of circum-galactic structure where cool clumps embed within a hot and highly-ionized halo appears to change (Becker et al. 2011b; Stern et al 2016), as the signature of the high-ionization phase diminishes (Figure 4). This may result from a softening of the ionizing background spectrum (Cooper et al. 2019), possibly during the late stages of reionization.

To date, the highest-redshift absorbers show anomalies in ionization, but not in abundances; the inferred element yields (e.g, α/Fe ratios) remain consistent with Pop II stars (Becker et al. 2012). These measurements can be straightforwardly extended to epochs wherever QSOs can be observed, projected presently as early as z = 8–9 (Fan et al. Astro2020 white paper), where the nucleosynthetic signatures of Pop III stars may be detected.

<u>Future prospects</u>: Absorption science in the next decade requires access to ELT-scale apertures with spectrometers covering the Y/J bands (for O I, C II, Si II, C I, Si IV, Al II) to the K band (for Mg II and Fe II), with R > 5000 to effectively split the OH sky background and provide a proper match in resolution to systems of interest (i.e. containing multiple components; R > 20,000 is required to resolve the individual components). Adaptive optics feeds are not required since QSOs do not reveal intrinsic morphology; however they will increase sensitivity for point-source observations in the presence of strong sky foregrounds, and potentially reduce the size and cost of instruments if the AO system is considered part of the telescope infrastructure. Notably, *JWST* and ALMA will play a complementary role in searches for galaxies near well-studied QSO sightlines. Such surveys are already planned as early GTO programs.

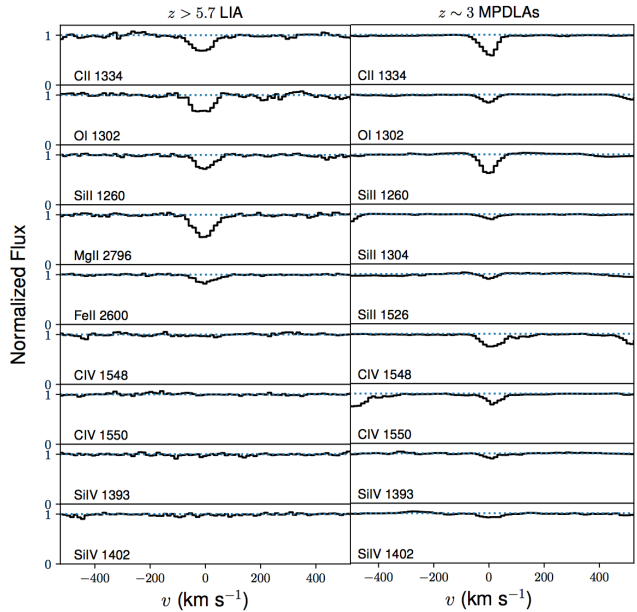

**Figure 4.** Stack of transitions from low-ionization absorption systems at z > 6 (left) and metal-poor damped Lyman alpha absorbers at z ~ 3 (right). The neutral species of C II, Si II, O I, and Mg II present similar absorption strengths, while high-ionization C IV and S IV transitions are much weaker at high redshift. The weakening of highly ionized gas likely reflects a softening of the ionizing UV background approaching reionization.

## IV. Summary

QSO absorption-line studies with the next generation of telescopes promise to help revolutionize our understanding of the reionization epoch. The science goals above require moderate- to high-resolution optical and near-IR spectrographs on ELTs, along with well-calibrated near-IR spectra from *JWST*. To understand early black hole growth, infrared all-sky surveys will devote major resources to searches for z > 7 quasars in the coming decade, and those efforts will have collateral benefits for this work. Because the total number of spectroscopic QSOs on the sky at z > 8 may only be of order N = 1–10, full-sky (northern and southern) ELT visibility is an important priority.

In closing, we note a parallel to the advent of 6-10m telescopes outfitted with efficient spectrographs and detectors, which transformed QSO absorption lines into a precision tool for astrophysics and cosmology out to z ~ 5. The next 10 years will push these advances into the new high-redshift frontier.